\documentclass[11pt,showpacs,preprintnumbers,superscriptaddress,amsmath,amssymb,nofootinbib]{revtex4}
\usepackage{graphicx}% Include figure files
\usepackage{dcolumn}% Align table columns on decimal point
\usepackage{bm}% bold math
\usepackage{amssymb}
\usepackage{amsmath}
\usepackage{epsfig}    
\usepackage{color}
\usepackage{hhline}
\def\be{\begin{equation}}
\def\ee{\end{equation}}
\newcommand{\bea}{\begin{eqnarray}}
\newcommand{\eea}{\end{eqnarray}}
\newcommand{\nn}{\nonumber}

\numberwithin{equation}{section}

\begin{document}

%%%%%%%%
%\title{Radiative Neutrino Mass in Alternative Left-Right Symmetric Model}

\title{Radiative Neutrino Mass  in  Alternative Left-Right  Model}

\preprint{KIAS-P16020}
\author{Takaaki Nomura}
\email{nomura@kias.re.kr}
\affiliation{School of Physics, KIAS, Seoul 130-722, Korea}

\author{Hiroshi Okada}
\email{macokada3hiroshi@cts.nthu.edu.tw}
\affiliation{Physics Division, National Center for Theoretical Sciences, Hsinchu, Taiwan 300}

\author{Yuta Orikasa}
\email{orikasa@kias.re.kr}
\affiliation{School of Physics, KIAS, Seoul 130-722, Korea}
\affiliation{Department of Physics and Astronomy, Seoul National University, Seoul 151-742, Korea}

\date{\today}

\begin{abstract}
We propose a radiative seesaw model in {alternative left-right model} without any bidoublet scalar fields, in which
all the fermion masses in the  standard  model are generated through {a canonical seesaw mechanism at the tree level}.
On the other hand the observed neutrino masses are generated at two-loop level.
In this paper we focus on the neutrino sector and show how to induce the active neutrino masses.
Then we discuss the observed neutrino oscillation, constraints from lepton flavor violations, new sources of muon anomalous magnetic moment, a long-lived dark matter candidate with keV scale mass, and collider physics.
\end{abstract}
\maketitle
\newpage

\section{Introduction}
Current neutrino oscillation data {provide} strong evidence of {tiny  but nonzero neutrino masses}~\cite{Forero:2014bxa}.
Seesaw mechanism is one of the elegant realization to explain such {tiny neutrino masses} by introducing right-handed neutrinos, which 
can naturally be embedded into a left-right symmetry {$SU(2)_L \times SU(2)_R \times U(1)_{B-L}$} as a theory at TeV scale~\cite{Mohapatra:1974hk}.~\footnote{The left-right symmetry can smoothly be extended into  larger groups such as $SO(10)$ symmetry, which is typically realized at higher scale such as grand unified theories.}

On the other hand, radiative seesaw models are also {one of} the natural realizations to explain the tiny neutrino masses at low energy scale where the neutrino mass matrix is generated at loop level, and a vast paper has recently been arisen in Refs.~\cite{Zee, Cheng-Li, zee-babu, Krauss:2002px, Ma:2006km, Aoki:2008av, Gustafsson:2012vj, Hambye:2006zn, Gu:2007ug, FileviezPerez:2016erl, Sahu:2008aw, Gu:2008zf, Babu:2002uu, AristizabalSierra:2006ri, AristizabalSierra:2006gb,
Nebot:2007bc, Bouchand:2012dx, Kajiyama:2013sza,McDonald:2013hsa, Ma:2014cfa, Schmidt:2014zoa, Herrero-Garcia:2014hfa,
Ahriche:2014xra,Long1, Long2, Aoki:2010ib, Kanemura:2011vm, Lindner:2011it,
Kanemura:2011jj, Aoki:2011he, Kanemura:2011mw, Schmidt:2012yg, Kanemura:2012rj, Farzan:2012sa, Kumericki:2012bf, Kumericki:2012bh, Ma:2012if, Gil:2012ya, Okada:2012np, Hehn:2012kz, Baek:2012ub, Dev:2012sg, Kajiyama:2012xg, Kohda:2012sr, Aoki:2013gzs, Kajiyama:2013zla, Kajiyama:2013rla, Kanemura:2013qva,Law:2013saa, Dasgupta:2013cwa, Baek:2013fsa, Baek:2014qwa, Okada:2014vla, Ahriche:2014cda, Ahriche:2014oda,Chen:2014ska,
Kanemura:2014rpa, Okada:2014oda, Fraser:2014yha, Okada:2014qsa, Hatanaka:2014tba, Baek:2015mna, Jin:2015cla,
Culjak:2015qja, Okada:2015nga, Geng:2015sza, Okada:2015bxa, Geng:2015coa, Ahriche:2015wha, Restrepo:2015ura, Kashiwase:2015pra, Nishiwaki:2015iqa, Wang:2015saa, Okada:2015hia, Ahriche:2015loa, Ahn:2012cg, Ma:2012ez, Kajiyama:2013lja, Hernandez:2013dta, Ma:2014eka, Aoki:2014cja, Ma:2014yka, Ma:2015pma, Ma:2013mga,
%%%
radlepton1, radlepton2, Okada:2014nsa, Brdar:2013iea, Okada:2015nca, 
%%%
Okada:2015kkj, Fraser:2015mhb, Fraser:2015zed, Adhikari:2015woo, Kanemura:2015cca, Bonnet:2012kz,Sierra:2014rxa, Davoudiasl:2014pya, Lindner:2014oea,Okada:2014nea, MarchRussell:2009aq, King:2014uha, Mambrini:2015sia, Boucenna:2014zba, Ahriche:2016acx, Okada:2015vwh, Ahriche:2016rgf, Kanemura:2015bli, Nomura:2016fzs, Yu:2016lof, Ding:2016ldt, Nomura:2016seu, Okada:2016rav, Nomura:2016rjf, Ko:2016sxg, Hernandez:2015hrt, Arbelaez:2016mhg, Chao:2015nac}. 
%%%
Moreover some new particles {such as dark matter (DM) and/or electrically charged particles}, running inside a loop diagram, are introduced in radiative seesaw models. Thus the radiative seesaw models provide a wide variety of interesting phenomenologies correlated with neutrino sector,
{and these two scenarios are well compatible~\cite{Gu:2008zf, FileviezPerez:2016erl}.
Thus it is an attractive interpretation that the active neutrino masses are generated by combination of these mechanisms since 
neutrino masses are very light compared to the other standard model (SM) fermions.
In addition, implementing this scenario into left-right model will be phenomenologically interesting.}

In this paper, we combine the left-right model and radiative seesaw model, in which 
active neutrino masses are generated at two loop level while Dirac neutrino masses are generated at one loop,
%in which the {Dirac} neutrino masses are generated at the one loop level 
%%%
employing a specific left-right model based on Ref.~\cite{Mohapatra:1987nx}~\footnote{The paper also discusses the quark sector.}.
And a Majorana mass term of right-handed neutrino is obtained at tree level by introducing $SU(2)_R$ triplet scalar $\Delta_R$.
But we do not assume the exact left-right symmetry and $\Delta_L$ is not introduced.
Then we find allowed region of parameter spaces by carrying out numerical analysis where we take into account muon anomalous magnetic moment, various lepton flavor violating processes, and a long lived DM candidate to explain the x-ray line at 7.1 keV~\cite{Boyarsky:2014jta, Bulbul:2014sua}, as well as consistency with the current neutrino oscillation data.

%in which an isospin singlet singly charged boson $h^\pm$, doubly charged boson $k^{\pm\pm}$, and singly fermion are introduced.

This paper is organized as follows.
In Sec.~II, we show our model building including Higgs masses, neutrino mass matrix.
In Sec.~III, we discuss {lepton flavor violations (LFV)}, muon anomalous magnetic  moment,  DM, and collider physics and then carry out numerical analysis to search for the parameter space satisfying all the phenomenological constraints.
We conclude in Sec.~VI.

%%%%%%%%%%%%%%%%%%%%%%%%%%%%%%%%%%%%%%
\section{The Model}

In this section, we introduce our model where the gauge symmetry is introduced as $SU(2)_L \times SU(2)_R \times U(1)_{B-L}$.
In this paper, we focus on the lepton sector and the details of the quark sector is found in Ref.~\cite{Mohapatra:1987nx}.

\subsection{Particle contents and scalar sector}
%
%\subsection{Model setup}
\begin{table}[t]
\centering {\fontsize{10}{12}
\begin{tabular}{|c||c|c|c|}
\hline Fermion & $L_L$ & $ L_{R} $ & $E_{L(R)}$  %& $e'_{L(R)}$  
  \\\hhline{|=#=|=|=|$}
$(SU(2)_L,SU(2)_R,U(1)_{\rm B-L})$ & $(\bm{2},\bm{1},-1)$ & $(\bm{1},\bm{2},-1)$ & $(\bm{1},\bm{1},-2)$   %& $(\bm{1},-1)$
\\\hline
%$U(1)_L$ & $+1$ & $-1$  & $-1$ & $+1$ & $-1$ & $0$ & $0$  & $0$  \\\hline
%%%
%$U(1)'$ & $-1/2$ & $-1/2$ & $-1/2$ & $-1/2$ & $3/2$  \\\hline
%$U(1)'$ & $\ell$ & $\ell$ & $\ell$ & $\ell$ & $\ell-\eta_2$ & $\ell-\eta_2$  \\\hline
%$U(1)_{} $ & $-1$ & $-1$ &  $-3/2$ %&  $-3/2$    \\\hline
%$Z_2$ & $+$ & $+$ &  $-$  \\\hline
%%%
\end{tabular}%
} \caption{Lepton sector; notice the three  flavor index of each field $L_{L(R)}$ and $E_{L(R)}$ is abbreviated.} 
\label{tab:1}
\end{table}
\begin{table}[t]
\centering {\fontsize{10}{12}
\begin{tabular}{|c||c|c|c|c|}
\hline Boson  & $\Phi_L$   & $\Phi_R$    & $h^+$  & $\Delta_R$    　%& $\varphi$   & $k^{++}$
  \\\hhline{|=#=|=|=|=|}
$(SU(2)_L,SU(2)_R,U(1)_{\rm B-L})$ & $(\bm{2},\bm{1},1)$  & $(\bm{1},\bm{2},1)$   & $(\bm{1},\bm{1},2)$  
& $(\bm{1},\bm{3},2)$    \\\hline %   & $(\bm{1},\bm{1},4)$
%$U(1)_L$ & $0$ & $0$ & $0$  & $0$  \\\hline
%%%
%$U(1)' $  & $0$ & $0$ & $1$  & $-1$   & $2$  \\\hline
%$U(1)' $  & $0$ & $0$ & $\eta_2$  & $\eta_2-2\ell$   & $-2\ell$  \\\hline
%$U(1)_{} $ & $0$ & $3/2$ &  $1$ & $3$ &  $1/2$    \\\hline
%$Z_2$ & $+$ & $-$ &  $+$  \\\hline
%%%
\end{tabular}%
} 
\caption{Boson sector }
\label{tab:2}
\end{table}

The particle contents for leptons and bosons are respectively shown in Tab.~\ref{tab:1} and Tab.~\ref{tab:2}.
Here all the new fields are singlet under $SU(3)_C$.
We introduce {$SU(2)_R$} doublet  fermions of $L_{R}$ and isospin singlet vector-like fermions of $E_{L(R)}$ both of which have three flavors like SM fermions. % to the SM fields.
%Each of the exotic field needs (at least) two flavors in order to satisfy current neutrino oscillation data~\cite{pdf}. 
%%%
%{\color{red} Note that the particle contents in the quark sector is the same as Ref.~\cite{Mohapatra:1987nx} and we omit to show them here.}
As for new bosons, we introduce
two {$SU(2)_{L(R)}$} doublet scalars $\Phi_L$  and $\Phi_R$,
an isospin singlet singly-charged scalar $h^{\pm}$, and {an} $SU(2)_R$ triplet scalar $\Delta_R$.
%and an  isospin singlet neutral scalar $\varphi$.
Note here that $\Phi_R$ and $\Delta_R$ {respectively} develop {vacuum expectation values (VEVs)} (denoted by $v_R/\sqrt2$ {and $v_{\Delta}/\sqrt2$}) in order to break the $SU(2)_R$ symmetry
{and generate Majorana mass term for the right-handed neutrinos $\nu_R$ to realize seesaw mechanism with two-loop induced Dirac mass as shown below.}
%{Here the singlet charged scalar $h^\pm$ is introduced to explore Zee-Babu type Majorana mass for active neutrinos which will be compared with the Majorana mass from the seesaw mechanism.  }
%$\varphi$, (which is assumed to be the parity odd singlet), plays an role in obtaining the desired breaking pattern of $\Phi_L$ and $\Phi_R$~\cite{moha}.

The relevant  Lagrangian for Yukawa sector and scalar potential under these assignments
are given by
\begin{align}
-\mathcal{L}_{Y}
=&
(h_{L})_{ij} \bar L_{L_i} \Phi_L E_{R_j} + (h_{R})_{ij} \bar L_{R_i} \Phi_R E_{L_j}
+(f_L)_{ij} \bar L_{L_i}^C i\tau_2 L_{L_j} h^+ + (f_R)_{ij} \bar L_{R_i}^C i\tau_2 L_{R_j} h^+\nn\\
& %+ g_{ij} \bar E^C_i E_j k^{++} 
+( y_{\Delta_R})_{i} \bar L^c_{R_i} i\tau_2 \Delta_R L_{R_i}
+ (M_{E})_{i} \bar E_i E_i %+i  y_{ij}\varphi \bar E_i\gamma_5 E_j 
+{\rm c.c.},
 \label{Lag:Yukawa}
%%%
\end{align}
\begin{align}
\mathcal{V}
&=
 -m^2_{\Phi_L} |\Phi_L|^2 -m^2_{\Phi_R} |\Phi_R|^2  -m^2_{h} |h^+|^2   -m^2_{\Delta} {\rm Tr}[|\Delta_R|^2]
 %\nn\\- m^2_\varphi |\varphi|^2 \nn\\&+\mu_L (\varphi |\Phi_L|^2 +{\rm h.c.}) +\mu_R (\varphi |\Phi_R|^2 +{\rm h.c.})&
%+\frac{\mu_1}2 (h^- h^- k^{++} +{\rm h.c.}) 
+\frac{\mu_2}2 (\Phi_R^T i \tau_2\Delta_R^\dagger \Phi_R  +{\rm h.c.})
  \nn\\
&
+\lambda_{\Phi_L}|\Phi_L|^4 + \lambda_{\Phi_R}|\Phi_R|^4 + \lambda_{h}|h^+|^4 %+ \lambda_{k}|k^+|^4   
+ \lambda_{\Delta}({\rm Tr}[|\Delta_R|^2])^2 + \lambda'_\Delta {\rm Tr}[|\Delta_R|^4]\nn\\
%+ \lambda_{\varphi}|\varphi|^4 
& +\lambda_{{LR}}|\Phi_L|^2|\Phi_R|^2
%\nn\\&
+\lambda_{{Lh}}|\Phi_L|^2|h^+|^2 
% +\lambda_{{Lk}}|\Phi_L|^2|k^{++}|^2 +\lambda_{{L\varphi}}|\Phi_L|^2|\varphi|^2
+\lambda_{{Rh}}|\Phi_R|^2|h^+|^2  
\nn\\&
% +\lambda_{{Rk}}|\Phi_R|^2|k^{++}|^2 +\lambda_{{R\varphi}}|\Phi_R|^2|\varphi|^2+\lambda_{{hk}}|h^+|^2|k^{++}|^2
 +\lambda_{h\Delta}|h^+|^2{\rm Tr}[|\Delta_R|^2]  
%+\lambda_{k\Delta}|k^{++}|^2{\rm Tr}[|\Delta_R|^2]
 %\nn\\
 % \nn\\&+\lambda_{{h\varphi}}|h^+|^2 |\varphi|^2 +\lambda_{{k\varphi}}|k^{++}|^2 |\varphi|^2 &
  +\lambda_{\Phi_L\Delta}|\Phi_L|^2{\rm Tr}[|\Delta_R|^2]  +\lambda_{\Phi_R\Delta}|\Phi_R|^2{\rm Tr}[|\Delta_R|^2]
%\lambda'_{\Phi_L\Delta}\sum_{i}^{1-3}(\Phi^\dag_L \tau_i\Phi_L){\rm Tr}[\Delta^\dag_R \tau_i \Delta_R] \nn\\&
  + \lambda'_{\Phi_R\Delta} \Phi_R^\dagger \Delta_{{R}} \Delta_{ {R}}^\dagger \Phi_R 
 %+ \lambda_{\phi_R \Delta k} (\Phi_R^T i \tau_2 \Delta_R \Phi_R k^{--} + h.c)
%\lambda'_{\Phi_R\Delta}\sum_{i}^{1-3}(\Phi^\dag_R \tau_i\Phi_R){\rm Tr}[\Delta^\dag_R \tau_i \Delta_R]
,
\label{HP}
\end{align}
where $\tau_2$ is a second component of the Pauli matrix, the index $i(j)$ runs $1$-$3$, and $y_{\Delta_R}$ and $M_E$ can be diagonal without loss of the generality. It implies that $y_{\Delta_R}$ does not contribute to  lepton flavor violations. 
Notice here that each of $f_{L(R)}$ and $g$ should be anti-symmetric and symmetric.
We work on the basis where all the coefficients are real and positive for our brevity. 
%The couplings $\lambda_1$, $(\lambda_2)_{iiii}\ (i=1\,,2)$, $\lambda_{8}$, and  $\lambda_{9}$ have to be positive to stabilize the Higgs potential.
%%%
%\subsection{Scalar sector}y_\Delta
After the left-right symmetry breaking, each of scalar field has nonzero mass.
We parametrize  these scalar fields as 
\begin{align}
\Phi_{L(R)} &=\left[
\begin{array}{c}
\phi^+_{L(R)}\\
\phi^0_{L(R)}
\end{array}\right],\
\quad
\phi^0_{L(R)}=\frac1{\sqrt2}(v_{L(R)} + h_{L(R)} + ia_{L(R)}),\\
%%%
\Delta_R& =\left[
\begin{array}{cc}
\frac{\Delta^+}{\sqrt2} & \Delta^{++}\\
\Delta^0 & -\frac{\Delta^+}{\sqrt2}
\end{array}\right],\quad
\Delta^0 = \frac1{\sqrt2}(v_{\Delta} + \Delta_R^{ {0}} + i\Delta_I^{ {0}}),
\label{component}
\end{align}
where $h_L$ is the SM-like Higgs, and $v_L$ is related to the Fermi constant $G_F$ by $v_L^2=1/(\sqrt{2}G_F)\approx(246$ GeV)$^2$.
%$\phi^\pm_{L(R)}$ and $a_{L(R)}$ are respectively absorbed by the longitudinal component of the extra bosons $W_{L(R)}^\pm$ and $Z^0_{L(R)}$ associated with the left-right symmetry.
The VEVs of $\Phi_{L(R)}$ are derived from the conditions $\partial \mathcal{V}/ \partial v_L =0$, $\partial \mathcal{V}/ \partial v_R =0$ and $\partial{V}/\partial v_\Delta =0$ such that
\begin{align}
\label{eq:VEV}
v_L^2 & \simeq \frac{1}{\lambda_{\Phi_L}} \left( m_{\Phi_L}^2 - \frac{\lambda_{LR}}{2} v_R^2 \right), \nonumber \\
v_R^2 & \simeq \frac{1}{\lambda_{\Phi_R}} \left( m_{\Phi_R}^2 - \frac{\lambda_{LR}}{2} v_L^2 \right),\\
v_\Delta^2 & \simeq - \frac{1}{2 \sqrt{2}} \frac{\mu_2 v_R^2}{m_\Delta^2}.
\end{align}
In this paper we require $v_R \gg v_L $ which can be achieved if we adopt $m_{\Phi_R}^2/\lambda_{\Phi_R} \gg m_{\Phi_L}^2/\lambda_{\Phi_L}$ and {choose rather small value of $\lambda_{LR}$}.
After the symmetry breaking, we have massive gauge bosons $W_{L(R)}^\pm$ and $Z_{L(R)}$ associated with left-right symmetry.
{Note that neutral singlet scalar is required to obtain desired symmetry breaking pattern in the model of Ref.~\cite{Mohapatra:1987nx} while we can realize the symmetry breaking due to the absence of exact left-right symmetry in the scalar potential.}

%$\eta$ is the inert doublet and the mass of $\eta$ should be positive. In our model, $\eta$ mass is generated by the mixing between $\eta$ and $\varphi$. Consequently, the mixing should be positive at the symmetry breaking scale, 
%\begin{eqnarray}\lambda_{\eta\varphi}>0. \label{inert1}\end{eqnarray}
%And the quartic couplings satisfy the following inert conditions\cite{Barbieri:2006dq}, 
%\begin{eqnarray}\lambda_\Phi>0,\  \lambda_\eta>0,\  
%\lambda_{\Phi\eta}+\lambda'_{\Phi\eta}-\mid\lambda''_{\Phi\eta}\mid>-2\sqrt{\lambda_\Phi\lambda_\eta}. \label{inert2}\end{eqnarray}

The CP even Higgs boson mass matrix in the basis of ($h_L, h_R,\Delta_R^{0}$) is denoted by
$(M^{2})_{\rm CP-even}$, and it is diagonalized by 3 $\times$ 3 orthogonal mixing matrix $O_R$ as
$O_R (M^{2})_{\rm CP-even} O_R^T=$ diag.$(m_{h_{1}}^2,m_{h_2}^2,m_{h_3}^2)$.
{Thus $h_{L(R)}$ and $\Delta_R^0$ are respectively given by 
\begin{equation}
h_L\equiv \sum_{a=1-3} (O_R^T)_{1a}h_a, \qquad
h_R\equiv \sum_{a=1-3} (O_R^T)_{2a}h_a,\qquad
\Delta_R^0 \equiv \sum_{a=1-3} (O_R^T)_{3a}h_a,
\end{equation}
where $h_1\equiv h_{\rm SM}$ is the SM Higgs and $h_{2,3}$ are additional CP even Higgs mass eigenstates. }
 
 The CP odd component $a_L$ from $\Phi_L$ does not mix with the other CP odd components.
 Thus $a_L$ is the massless {Nombu-Goldstone (NG)} boson which is absorbed {by} $Z_L$ boson.
 The CP odd Higgs boson mass matrix in the basis of ($\Delta_I^{ {0}}, a_R^0$) is denoted by
$(M^{2})_{\rm CP-odd}$, and it is diagonalized by 2 $\times$ 2 orthogonal mixing matrix $O_I$ as
$O_I (M^{2})_{\rm CP-odd} O_I^T=$ diag.$(m_{A_{1}}^2,m_{A_2}^2)$.
{Therefore $a_R$ and $\Delta_I$ are given by 
\begin{equation}
a_R\equiv \sum_{a=1-2} (O_I^T)_{1a}A_a, \qquad  \Delta_I \equiv \sum_{a=1-2} (O_I^T)_{2a}A_a,
\end{equation}
where only $A_1$ should be massive, since $A_2$ is absorbed {by} $Z_R$ boson. }
 
The singly charged scalar boson $h^+$ does not mix with other charged scalar bosons. Thus it is the mass eigenstate with mass $m_{h^\pm}$.
Also the singly charged component $\phi^\pm_{L}$ from $\Phi_L$ does not mix and it is the massless NG boson absorbed by $W_L^\pm$.
The  singly charged scalar boson mass matrix in the basis of $(\Delta^\pm,\phi_R^\pm)$ is denoted by
$(M^{2})_{\rm singly}$, and it is diagonalized by 2 $\times$ 2 unitary mixing matrix $U_1$ as
$U_1 (M^{2})_{\rm singly} U_1^\dag=$ diag.$(m_{\phi^\pm_{1}}^2,m_{\phi^\pm_2}^2)$.
{Therefore $\Delta^\pm$ and $\phi^\pm_R$ are given by
\begin{equation}
\Delta^\pm\equiv \sum_{a=1-2} (U_1^\dag)_{1a}\phi^\pm_a, \qquad
\phi_R^\pm\equiv \sum_{a=1-2} (U_1^\dag)_{2a}\phi^\pm_a,
\end{equation}
where $m_{\phi^\pm_2}^2$  should be zero, since $\phi^{\pm}_2$ is absorbed {by} $W^\pm_R$ boson.  }
The  doubly charged scalar boson $\Delta^{\pm\pm}$  is mass eigenstate with mass eigenvalue $m_{\Delta^{\pm \pm}} \simeq m_\Delta$. 
%denoted by$(M^{2})_{\rm doubly}$, and it is diagonalized by 2 $\times$ 2 unitary mixing matrix $U_{2}$ as
% $U_{2} (M^{2})_{\rm doubly} U_{2}^\dag=$ diag.$(m_{h^{++}_{1}}^2,m_{h^{++}_2}^2)$.
%{Therefore $k^{\pm \pm}$ and $\Delta^{\pm \pm}$ are given by 
%\begin{equation}
%k^{\pm\pm} \equiv \sum_{a=1-2} (U_{2}^\dag)_{1a}h^{\pm\pm}_a,  \qquad  \Delta^{\pm\pm}\equiv \sum_{a=1-2} (U_{2}^\dag)_{2a}h^{\pm\pm}_a.\end{equation}}
%\newpage
%\subsection{Lepton sector}
%In this subsection, we will discuss the charged lepton sector and neutrino sector, respectively.

\subsection{Charged lepton sector}
First of all, we define the isospin doublet fermions as $L_{L(R)}\equiv [\nu_{L(R)},\ell_{L(R)}]^T$.
The charged lepton mass matrix in the basis of $(\ell, E)$ can be given as
\begin{align}
M_\ell =\left[
\begin{array}{cc}
0 & h_L v_L/\sqrt2\\
h_R^T v_R/\sqrt2 & M_E\\
\end{array}\right]
\equiv 
\left[
\begin{array}{cc}
0 & m_L\\
m_R^T & M_E\\
\end{array}\right].
\end{align}
Then it can be diagonalized by bi-unitary mixing matrix $V_L$ and $V_R$ as $V_L M_\ell V^T_R=M_{\rm diag}$, where
\begin{align}
 V_L M_\ell V^T_R &\approx
\left[
\begin{array}{cc}
- m_L M^{-1}_E m_R & 0\\
0& M_E \\
\end{array}\right],
\\
V_a &\approx
\left[
\begin{array}{cc}
1 - \frac{\rho_a \rho_a^T}{2} &-\rho_a\\
\rho_a^T & 1 - \frac{\rho_a^T \rho_a}{2}  \\
\end{array}\right], \quad \rho_L=m_L M^{-1}_E,\quad \rho_R =m_R^T M^{-1}_E ,\quad { a=L,R}.
\end{align}
Here we have used the assumption $m_L,m_R << M_E$.
The resultant charged lepton mass squared is then given by
\begin{align}
|m_\ell |^2_{ij}
\approx 
m_L M_E^{-1}m_R m_R^\dag (M_E^{-1}) m_L^\dag
= \frac{v^2_L v^2_R}{4} h_L M^{-1}_E h_R h_R^\dag M^{-1}_E h_L^\dag \approx \frac{v^2_L}{4} |h_L h_R|^2,
\end{align} 
if we assume to be $M_E\approx v_R$.

\subsection{Neutrino sector}
The neutral fermion mass matrix in the basis of $(\nu_L,\nu_R)$ is generated by% follows~\cite{Ma:2006km, Hehn:2012kz}:
\begin{align}
({\cal M}_\nu)_{ab}=
\left[
\begin{array}{cc}
0 & m_D\\
m_D^T & m_{\nu_R}\\
\end{array}\right],
 \label{eq:neut-theory}
\end{align}
%%%
where $m_{\nu_R}\equiv y_{\Delta_R}v_\Delta{/\sqrt2}={\rm diag.}(m_{N_1},m_{N_2},m_{N_3})$~
%%%%%%
\footnote{Our main motivation to introduce the $SU(2)_R$ triplet boson $\Delta_R$ is to formulate the seesaw neutrino mass matrix appropriately. Actually even if $\Delta_R$  is not introduced, rather heavier right-handed neutrino mass matrix $m_{\nu_R}$ can be induced at the two-loop level by increasing the scale of $v_R$.
However {we cannot define its inverse of the seesaw neutrino mass matrix, because the matrix rank  is reduced by one. Therefore, the seesaw formula does not work well.}
},
%%%%%%%
and {the Dirac fermion mass matrix $m_D$} is given by
%\textcolor{blue}
{
\begin{align}
%m_{\nu_{L}}&=\frac{\mu_1 v_{L}^2 (F_{L})^\rho_{i\alpha} (h_{L})^{}_{\alpha a} G^{*\sigma}_{ab} (h_{L}^{ T})_{b\beta} (F_{L}^{T})^\omega_{\beta j} }
%\frac{\mu v_{L}^2 (F_{L})^\rho_{i\alpha} (H_{L})^{\delta}_{\alpha a} G^{*\sigma}_{ab} (H_{L}^{ T})^\gamma_{b\beta} (F_{L}^{T})^\omega_{\beta j} }
%{32\pi^4 M^2_{\rm max}} {\rm F_2},\\
%%%
%{\rm F_2}&=\int \frac{dxdydz d\alpha d\beta d\gamma\delta(x+y+z-1)\delta(\alpha +\beta +\gamma-1)}
%{(z^2-z)(\alpha X_{E_a}+\beta X_{h^\pm})-\gamma(x X_{E_b}+y X_{h^\pm}+z X_{k^{\pm\pm}})},
%\\
%%%
m_D&\approx
\frac{v_{L} v_R (F_{L})^1_{i\alpha} (h_{L})_{\alpha a} (h_{R}^T)_{a\beta} (F_{R}^T)^1_{\beta j} }
%\frac{v_{L} v_R (F_{L})^\rho_{i\alpha} (H_{L})^\delta_{\alpha a} (H_{R}^T)^\gamma_{a\beta} (F_{R}^T)^\omega_{\beta j} }
{2\pi^2 M_{E_a}} \frac{\ln Z_{a,1}}{1-Z_{a,1}},
\label{eq:mD}
\end{align}
}
where all the indices are summed over, and we define $(F_{L/R})^a_{ij}\equiv (U_1^\dag)_{1a}(f_{L/R})_{ij}$,  
%$G^a_{ij}\equiv (U_{2}^\dag)_{1a}g_{ij}$, $(H_{L})^a_{ij}\equiv \sum_{a=1-3}(O_{L}^T)_{1a}(h_{L})_{ij}$, $(H_{R})^a_{ij}\equiv \sum_{a=1-3}(O_{R}^T)_{2a}(h_{R})_{ij}$, $X_{f}\equiv \left(\frac{m_{f} }{M_{\rm max}}\right)^2$, 
$Z_{a,\rho}\equiv \left(\frac{m_{h_\rho^{\pm}} }{M_{E_a}}\right)^2$, 
%$M_{\rm max}\equiv {\rm Max}[m_{h_\rho^\pm},m_{k_\sigma^{\pm\pm}},m_{h_\omega^\pm},M_{E_{a(b)}}]$, 
and assume to be $m_\ell << M_E$. 
%Here ${\rm F}_2$ should be the same formula as the Zee-Babu model~\cite{Herrero-Garcia:2014hfa} {except all the masses of  running fermions in the loop cannot be assumed to be massless}.
%%
%Supposing $v_L << v_R$, 
%We can take the following hierarchy $m_{\nu_L}<<m_D << m_{\nu_R}$ naturally, because these are respectively generated at two-loop, one-loop, and tree-level.
Therefore the active neutrino masses can be obtained at two-loop level  through two types of the seesaw mechanisms (canonical seesaw with one-loop induced Dirac mass and its irreducible diagram~\cite{Kanemura:2011mw});
$(\mathcal{M}_\nu)_{ab}\approx m_D m_{\nu_R}^{-1} m_D^T$.~\footnote{The loop function with the irreducible diagram is usually smaller than the one with the canonical seesaw diagram~\cite{Okada:2015nga}. Thus we consider the canonical seesaw type model only.}  
{Notice here that  one of three neutrino masses is zero without loss of the generality, because the matrix rank of $(m_D)_{3\times3}$ is two. }

%%%
%\textcolor{red}{
\if0
Here we estimate each of the order. First of all, we fix the loop function is the same order, therefore
\begin{align}
\left(\frac{\ln Z_{a,1}}{1-Z_{a,1}}\right)^2\approx F_2,
\end{align}
which is the appropriate assumption because these are rather independent of related mass parameters in the loop.
Then we derive the following value
\begin{align}
\frac{m_D m_{\nu_R}^{-1} m_D^T}{m_{\nu_L}}\approx
\frac{(v_R F_Rh_R M_{\rm Max})^2}{\mu_1 M_E^2 m_{\nu_R}}>>1,
\end{align}
where $[m_{\nu_R},\mu_1,M_E,v_R]<M_{\rm Max}$ and $F_R$ and $h_R$ are expected  not to be  so small
{due to originating from  the $SU(2)_R$.}
%, since these couplings originate from the $SU(2)_R$.
%%%
Thus we can take $(\mathcal{M}_\nu)_{ab}\approx m_D m_{\nu_R}^{-1} m_D^T$ {as a dominant contribution to the neutrino masses and mixings}, 
which is a natural assumption as far as the loop function is the same order among these loops and  there exists rather large hierarchy among mediating fields~\cite{Nomura:2016fzs}.
\fi
%}

%Remind here that two flavor of $E_k$(k=1-2) is introduced to obtain the current neutrino oscillation data.
Then $(\mathcal{M}_\nu)_{ab}$ {is} diagonalized by the Maki-Nakagawa-Sakata mixing matrix $V_{\rm MNS}$ (MNS) as
\begin{align}
(\mathcal{M}_\nu)_{ab} &=(V_{\rm MNS} D_\nu V_{\rm MNS}^T)_{ab},\quad D_\nu\equiv (m_{\nu_1},m_{\nu_2},m_{\nu_3}),
\\
V_{\rm MNS}&=
\left[\begin{array}{ccc} {c_{13}}c_{12} &c_{13}s_{12} & s_{13} e^{-i\delta}\\
 -c_{23}s_{12}-s_{23}s_{13}c_{12}e^{i\delta} & c_{23}c_{12}-s_{23}s_{13}s_{12}e^{i\delta} & s_{23}c_{13}\\
  s_{23}s_{12}-c_{23}s_{13}c_{12}e^{i\delta} & -s_{23}c_{12}-c_{23}s_{13}s_{12}e^{i\delta} & c_{23}c_{13}\\
  \end{array}
\right],
\end{align}
where we neglect {Dirac phase $\delta$ as well as Majorana phase}
in the numerical analysis for simplicity.
The following neutrino oscillation data at 95\% confidence level~\cite{pdf} is given as
\begin{eqnarray}
&& 0.2911 \leq s_{12}^2 \leq 0.3161, \; 
 0.5262 \leq s_{23}^2 \leq 0.5485, \;
 0.0223 \leq s_{13}^2 \leq 0.0246,  
  \\
&& 
%  m_{\nu_2} ({\rm eV}) = 0.0087,  \; 
  \ |m_{\nu_3}^2- m_{\nu_2}^2| =(2.44\pm0.06) \times10^{-3} \ {\rm eV}^2,  \; 
 % m_{\nu_3} ({\rm eV}) = 0.0502 .
  \ m_{\nu_2}^2- m_{\nu_1}^2 =(7.53\pm0.18) \times10^{-5} \ {\rm eV}^2, \nn
  \label{eq:neut-exp}
  \end{eqnarray}
where we assume {normal ordering of the neutrino mass eigenstate in our analysis below, therefore $m_{\nu_1}=0$.}

{
\subsection{Neutrinoless double beta decay}
Here we discuss the non-standard contribution to the neutrinoless double beta decay.
The relevant process arises from the same process of the standard interaction just by flipping the chirality $L\to R$, and its formula is given by
\begin{align}
m_{\beta\beta}&=
\left|
\sum_{i=1}^3|(V_{\rm MNS}^2)_{1i}|m_{\nu_i}+\left[\frac{m_{W_L}}{m_{W_R}}\right]^4\left[\frac{g_L}{g_R}\right]^4 m_{\nu_{R_1}}
\right|
=\left|
\sum_{i=1}^3|(V_{\rm MNS}^2)_{1i}|m_{\nu_i}+\frac{v_L^4}{(v_R^2+2 v^2_{\Delta})} m_{\nu_{R_1}}
\right|,
\end{align}
where the first term in the left side equation is the contribution to the SM and the second term is the one of the new contribution. 
Furthermore we have used $m_{W_L}=g_Lv_L/2$, $m_{W_R}=g_R\sqrt{v_R^2+2 v^2_{\Delta}}/2$, and the mixing among $\nu_R$s
is assumed to be diagonal for simplicity. When we adopt the typical bound $m_{\beta\beta}\lesssim0.29$ eV~\cite{KlapdorKleingrothaus:2006ff},
we can estimate the upper bound on the mass of $\nu_{R_1}$ once the $v_R$ and $v_{\Delta}$ are fixed.
We will see a concrete discussion in the section of numerical analysis.
}

%%%%%%%%%%%%%%%%%%%%%%%%%%%%%

\section{Phenomenology of the model}

{In this section, we discuss some phenomenologies in our model such as {LFV}, muon anomalous magnetic moment and {DM}.
Then numerical analysis is carried out to search for allowed parameter space which is consistent with current experimental data. }

\subsection{Muon anomalous magnetic moment and Lepton flavor violations}
%%%
{\it The muon anomalous magnetic moment} (muon $g-2$) has been 
measured at Brookhaven National Laboratory. 
The current average of  muon $g-2$ experimental results is found as~\cite{bennett}
\begin{align}
a^{\rm exp}_{\mu}=11 659 208.0(6.3)\times 10^{-10}. \notag
\end{align}
Two discrepancy between the
experimental data and the prediction in SM; $\Delta a_{\mu}\equiv a^{\rm exp}_{\mu}-a^{\rm SM}_{\mu}$,
have been respectively computed in Ref.~\cite{discrepancy1} as 
\begin{align}
\Delta a_{\mu}=(29.0 \pm 9.0)\times 10^{-10} \ {\rm at\ 3.2\sigma\ C.L.}, \label{dev1}
\end{align}
and in Ref.~\cite{discrepancy2} as
\begin{align}
\Delta a_{\mu}=(33.5 \pm 8.2)\times 10^{-10} \  {\rm at\ 4.1\sigma\ {C.L}}. \label{dev2}
\end{align}
%The above results given in Eqs. (\ref{dev1}) and (\ref{dev2}) correspond to $3.2\sigma$ and $4.1\sigma$ deviations, respectively. 

In our model, we have new contributions to $\Delta a_\mu$ coming from the Yukawa coupling of $h_{L(R)}$ and $f_{L(R)}$,
{and its} contribution is given as~\footnote{Useful formulae for the muon $g-2$ can be found in ref.~\cite{Queiroz:2014zfa}.}
\begin{align}
\Delta a_\mu &\approx \Delta a_\mu^{h_H} +  \Delta a_\mu^{h_A} + \Delta a_\mu^f + \Delta a_\mu^\Delta,\\
%%%
\Delta a_\mu^{h_H}  &\approx
\frac{m_\mu}{2(4\pi)^2}
\sum_{\alpha=1}^3 \sum_{a=1}^{3} 
\left[
%\frac{ }{  }
\frac{(H^a)_{2\alpha}(H^\dag_a)_{\alpha 2} }{M_{E_\alpha} }
\frac{1+3 Y_{\alpha,a}^2-4 Y_{\alpha,a}-2 Y_{\alpha,a}^2 \ln[Y_{\alpha,a}]}
{(1-Y_{\alpha,a})^3}\right], \\
%%%%%%%%%
\Delta a_\mu^{h_A}  &\approx
-\frac{1}{(4\pi)^2} \left(m_\mu\right)^2
\sum_{\alpha=1}^3 \sum_{a=1}^{3}  \frac{(H^{'a})^*_{\alpha,2} (H^{'a})_{\alpha,2}}{M_{E_\alpha}^2}
F_2\left(Y'_{\alpha,a}\right), \\
%%%%%%%%%
\Delta a_\mu^f &\approx
-
\frac{m_\mu^2}{3(4\pi)^2 m^2_{h^\pm} }
\sum_{\alpha,\beta=1}^3
\left[
(f^\dag_L)_{2\alpha}(f_L)_{\alpha2} + (f^\dag_R)_{2\beta}(f_R)_{\beta,2}F_2(\epsilon_\beta)
\right], \\
%%%%%%%%%
\Delta a_\mu^\Delta &\approx
-
\frac{m_\mu^2}{4(4\pi)^2 }
\sum_{\alpha,\beta=1}^3
\left[
\frac{(y^{\dag}_{\Delta_R} )_{2\alpha} (y_{\Delta_R})_{\alpha2}}{m^2_{\Delta^{\pm\pm}}} + 
\sum_{b=1}^{2} \frac{(Y^{b\dag}_{\Delta_1} )_{2\beta} (Y^{b}_{\Delta_1})_{\beta2}}{6 m^2_{h^{\pm}_b}} F_2(\epsilon^b_{\beta})
\right], \\
%\Delta a_\mu^\Delta &\approx-\frac{m_\mu^2}{4(4\pi)^2 }\sum_{\alpha,\beta=1}^3\sum_{a,b=1}^{2} 
%\left[\frac{(Y^{a\dag}_{\Delta_2} )_{2\alpha} (Y^{a}_{\Delta_2})_{\alpha2}}{m^2_{h^{\pm\pm}_a}} + \frac{(Y^{b\dag}_{\Delta_1} )_{2\beta} (Y^{b}_{\Delta_1})_{\beta2}}{6 m^2_{h^{\pm}_b}} F_2(\epsilon^b_{\beta})\right], \\
%%%%%%%%%
F_2(x)& \equiv 
\frac{1- 6x + 3 x^2 + 2x^3-6 x^2 \ln x}{(1-x)^4},
\label{eq:muon-g-2}
\end{align}
%%%%%%%
where we define $H^a_{ij}\equiv \frac{(h_L)_{ij} (O_R^T)_{1a}+ (h_R)_{ij} (O_R^T)_{2a} }{2\sqrt2}$, $H^{'a}_{ij}\equiv \frac{(O^T_I)_{1a}(h_R)_{ij}}{\sqrt2}$,  $Y_{\alpha,a}\equiv (m_{h_a^0}/M_{E_\alpha})^2$, $Y'_{\alpha,a}\equiv m^2_{A_a}/M^2_{E_\alpha}$, 
$\left(Y_{\Delta_1}\right)_{a \alpha}\equiv (U_1^\dag)_{2a} (y_{\Delta_R})_\alpha$, 
%$\left(Y_{\Delta_2}\right)_{a \alpha}\equiv (U_2^\dag)_{2a} (y_{\Delta_R})_\alpha$, 
$\epsilon_j^{(b)}\equiv (m_{\nu_{R_j}}/m_{h^\pm_{(b)}})^2$ and we have assumed $m_{{\nu_L}} << m_\ell<< \{ m_{{\nu_R}} , M_E, m_{h_a^0} ,m_{h^\pm_{(b)}} \}$.
Note here that the contribution of $\Delta a_\mu^\Delta$ is negligibly small because of the small mixing.

\begin{table}[t]
\begin{tabular}{c|c|c} \hline
Process & $(i,j)$ & Experimental bounds ($90\%$ CL) \\ \hline
%%%%%%%
$\mu^{-} \to e^{-} \gamma$ & $(2,1)$ &
	$\text{Br}(\mu \to e\gamma) < 5.7 \times 10^{-13}$  \\
$\tau^{-} \to e^{-} \gamma$ & $(3,1)$ &
	$\text{Br}(\tau \to e\gamma) < 3.3 \times 10^{-8}$ \\
$\tau^{-} \to \mu^{-} \gamma$ & $(3,2)$ &
	$\text{Br}(\tau \to \mu\gamma) < 4.4 \times 10^{-8}$  \\ \hline
\end{tabular}
\caption{Summary of $\ell_i \to \ell_j \gamma$ process and the lower bound of experimental data~\cite{Adam:2013mnn}.}
\label{tab:Cif}
\end{table}

\begin{table}[t]
\begin{tabular}{c|c|c} \hline
Process & $(i,j,k,\ell)$ & Experimental bounds ($90\%$ CL) \\ \hline
%%%%%%%
$\mu^{-} \to e^{-} e^{+} e^{-}$ & $(2,1,1,1)$ &
	$\text{Br}(\mu \to e^{-} e^{+} e^{-} ) < 1.0 \times 10^{-12}$  \\
$\tau^{-} \to e^{-} e^{+} e^{-}$ & $(3,1,1,1)$ &
	$\text{Br}(\tau \to e^{-} e^{+} e^{-} ) < 2.7 \times 10^{-8}$  \\
$\tau^{-} \to \mu^{-} e^{+} e^{-}$ & $(3,2,1,1)$ &
	$\text{Br}(\tau \to \mu^{-} e^{+} e^{-} ) < 1.8 \times 10^{-8}$  \\
$\tau^{-} \to e^{-} \mu^{+} \mu^{-}$ & $(3,1,2,1)$ &
	$\text{Br}(\tau \to e^{-} \mu^{+} e^{-} ) < 1.5 \times 10^{-8}$  \\
$\tau^{-} \to e^{-} \mu^{+} \mu^{-}$ & $(3,1,2,2)$ &
	$\text{Br}(\tau \to e^{-} \mu^{+} \mu^{-} ) < 2.7 \times 10^{-8}$  \\
$\tau^{-} \to \mu^{-} \mu^{+} \mu^{-}$ & $(3,2,2,2)$ &
	$\text{Br}(\tau \to \mu^{-} \mu^{+} \mu^{-} ) < 2.1 \times 10^{-8}$  \\
	\hline
\end{tabular}
\caption{Summary of $\ell_i^- \to \ell_j^- \ell_k^+, \ell_\ell^-$ process and the lower bound of experimental data~\cite{pdf}.}
\label{tab:3l}
\end{table}

%{\it  Lepton flavor violation processes} $\ell_i\to\ell_j\gamma$ and $\ell_i^-\to\ell_j^-\ell_k^+\ell^-_\ell$ at the one-loop level come from the same terms of anomalous magnetic moment at the one-loop level in principle. 
%%%
{{\it  Lepton flavor violation processes (LFVs)} $\ell_i\to\ell_j\gamma$ and $\ell_i^-\to\ell_j^-\ell_k^+\ell^-_\ell$ at the one-loop level 
are measured precisely and severely constrained.}
Each  of flavor dependent process has to satisfy the current upper bound, as can be seen in Table~\ref{tab:Cif} and \ref{tab:3l}.
The branching ratio (BR) for the $\ell_i\to\ell_j\gamma$ can be written as
\begin{align}
\text{Br}(\ell_i\to\ell_j\gamma) &= 
\frac{48 \pi^3\alpha_{\rm em} C_i }{m^2_{\ell_i} {\rm G_F}^2} \left(|a_L|^2+|a_R|^2\right),
\quad
a_L = a_{h} + a_{f_R},\quad a_R= a_{h} + a_{f_L}  ,\\
%%%%%
a_{h}
&\approx
-\frac{1}{2(4\pi)^2}
\sum_{\alpha=1}^3 
\sum_{a=1}^{3} 
\frac{(H^a)_{i\alpha}(H^\dag_a)_{\alpha j} }{M_{E_\alpha}}
\frac{1+3 Y_{\alpha,a}^2-4 Y_{\alpha,a}-2 Y_{\alpha,a}^2 \ln[Y_{\alpha,a}]}
{(1-Y_{\alpha,a})^3},\\
%%%%%
a_{f_R}
&\approx
m_{\ell_i} \sum_{\alpha=1}^3\frac{(f_R^\dag)_{j\alpha} (f_R)_{\alpha i}} {3(4\pi)^2m^2_{h^\pm}} F_2(\epsilon_\alpha),\quad
%\\%%%%%
a_{f_L}
\approx
m_{\ell_i}  \sum_{\alpha=1}^3\frac{(f_L^\dag)_{j\alpha} (f_L)_{\alpha i}} {3(4\pi)^2m^2_{h^\pm}},
 \label{eq:lfv-itoj}
\end{align}
where $C_i\approx(1,1/5)$ for $i=(\mu,\tau)$~\cite{Pich:2013lsa}, ${\rm G_F}$ is Fermi constant, and $\alpha_{\rm em}$ is the fine-structure constant.
On the other hand, the BR for the process $\ell_i^-\to\ell_j^-\ell_k^+\ell^-_\ell$ is given by
%%%
\begin{align}
&\text{Br}(\ell_i^-\to\ell_j^-\ell_k^+\ell^-_\ell)
\approx
\frac{C_i}{16  {\rm G_F}^2}
\left(8 |A|^2+ 8 |B|^2+2 |C|^2+2|D|^2
+2|A_L|^2 +2|B_L|^2 +2|A_R|^2 +2|B_R|^2\right.\nn\\
&\left.
+|C_R|^2
-8 {\rm Re}[AB^*] +4 {\rm Re}[AD^*]+4 {\rm Re}[BC^*]+{\rm Re}[CD^*]
-4{\rm Re}[AA_L^*] + 4{\rm Re}[AB_L^*] -4{\rm Re}[AA_R^*]\right.\nn\\
%%%%
&\left.
 + 4{\rm Re}[AB_R^*]+4{\rm Re}[A C_R^*] 
+4{\rm Re}[BA_L^*] - 4{\rm Re}[B B_L^*] + 4{\rm Re}[B A_R^*] - 4{\rm Re}[B B_R^*] -4{\rm Re}[B C_R^*]\right.\nn\\
%%%%
&\left.
 - 4{\rm Re}[A_L B_L^*] - 4{\rm Re}[A_R B_R^*] 
 - 4{\rm Re}[A_R C_R^*] + 4{\rm Re}[B_R C_R^*] + 8 {\rm Re}[B E^*] + 8 {\rm Re}[B F^*] + {\rm Re}[C E^*] \right. \nn\\
 %%%
&\left. + {\rm Re}[C F^*]
+ \frac12 |E|^2+ \frac12 |F|^2
  \right)
, \label{eq:lfv-itojkl}
\end{align}
%%%%
{where the numerical factors $\{A, B, C, D\}$ come from box loop diagrams in which $E_\alpha$ and $h_a$ are running while 
the other factors come from box loop diagrams with $E_\alpha$, $h_a$ and $A_a$ running inside the loop; the explicit forms of these factors are given in the Appendix B.
Note that the LFV decay ratios are determined by the Yukawa couplings $h_{L(R)}$ and $f_{L(R)}$ which also appear in our neutrino mass formula Eq.~(\ref{eq:mD}) 
indicating the correlation between LFV and neutrino mass matrix.}

\subsection{Dark Matter}
%%%%%%%%%%%%%%%%%%%
%%%%%%%%%%%%%%%%%%%
We consider a fermionic DM candidate $X(\equiv \nu_{R_1})$, which is assumed to be the lightest particle of $\nu_{R_{i}}$.
%However since $X$ decays into the SM particles ($\nu_L+\gamma$) at the one-loop level, heavier mass $\gtrsim{\cal O}$(1) GeV cannot be allowed due to its too fast decay. Hence
{Since} %However
DM can decay into neutrinos and photon through the Dirac mass term at the one-loop level, DM has to be long-lived. Hence we focus on the explanation of the X-ray line at 3.55 keV, since $X$ decays into active neutrinos and photon at the one-loop level after the 
{symmetry breaking.}
%spontaneous $U(1)$ symmetry breaking. 
Then the mass of DM $M_X(\equiv M_{\nu_{R_1}})$ is fixed to be around 7.1 keV with a small value of the decay rate divided by $M_X$;
{\it i.e.}, %\begin{align}
$4.8\times10^{-48}\lesssim\frac{\Gamma(X\to\nu_k\gamma)}{M_X}\lesssim4.6\times10^{-46}$~\cite{Faisel:2014gda}.
%\end{align}
\footnote{This bound is derived from $\sin^22\theta = (2-20)\times10^{-11}$.} 
{We also note that such a DM candidate will be over-abundant if one estimates thermal relic density through the gauge interactions.  
However this problem can be evaded by the entropy production due to the late decay of $\nu_{R_{2,3}}$~\cite{Bezrukov:2009th, Nemevsek:2012cd}.
In our analysis, we assume the right relic density can be obtained by this mechanism and the constraints on the decay rate of DM is taken into account.}
Then the decay rate is derived as
\begin{align}
& \frac{\Gamma(X\to\nu_k\gamma)}{M_X}\approx
\frac{\alpha_{\rm em} a b_j }{16\pi^4}
\left|
\sum_{j}^{1-3}{ (f_L^\dag)_{jk} (f_R)_{1j}} 
\frac{3-4 b_j+b_j^2+2 \ln[b_j]}{2(b_j-1)^3}
% \int dxdydx\delta(x+y+z-1) \frac{(1-x)^2}{x+(1-x)b_j }
\right|^2,
%\\&G\equiv \int_0^1dx \frac{(x-1)^3}{x^2-(Z_1-\epsilon_{j}-1) x+\epsilon_{j}(1-x) },
\label{eq:x-ray}
\end{align}
where we define $a\equiv (M_X/m_{h^\pm})^2$,  $b_j\equiv (m_{\ell_j}/m_{h^\pm})^2$,
under the assumption $M_X,m_{\nu_L}<<m_\ell,m_{h^\pm}$.
%and $\alpha_{\rm em}\approx1/137$ is the fine structure constant.
 %$\theta$ between $X$ and the active neutrinos; $\theta\approx5\times 10^{-6}$. 
{Thus the decay ratio is correlated to neutrino mass matrix, $\Delta a_\mu$ and $LFV$ through the Yukawa coupling $f_{L(R)}$.}

\subsection{Collider Physics}
%%%%%%%%%%%%%%%%%%
\begin{figure}[t]
\begin{center}
\includegraphics[width=70mm]{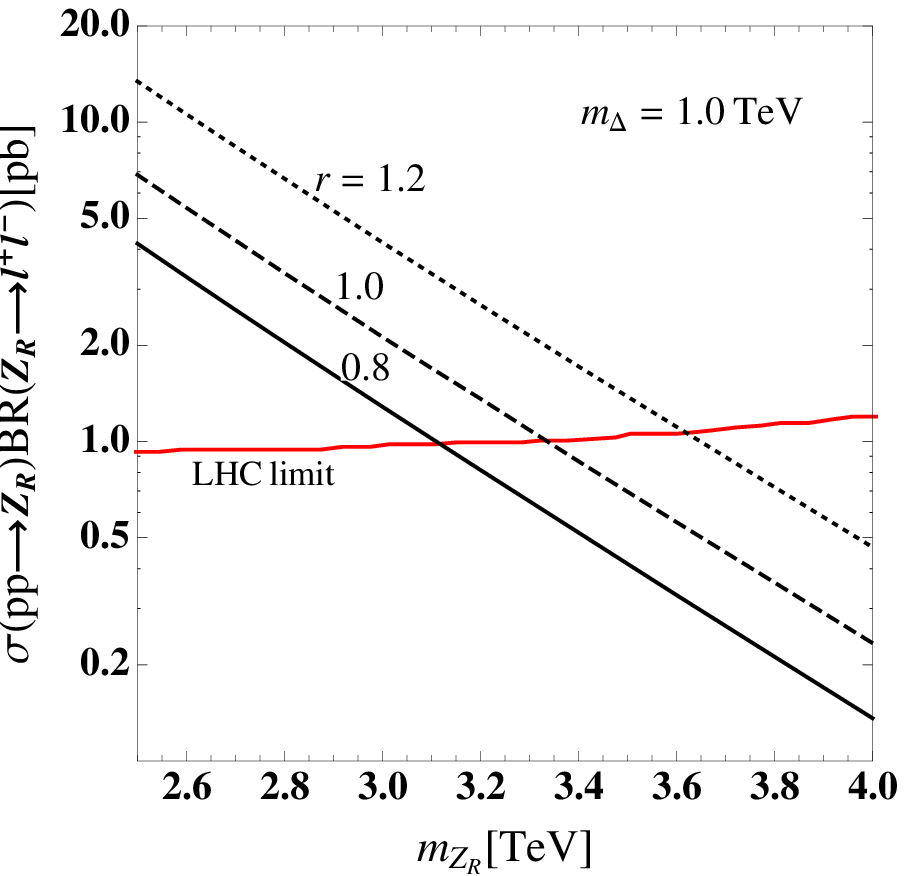}
\includegraphics[width=70mm]{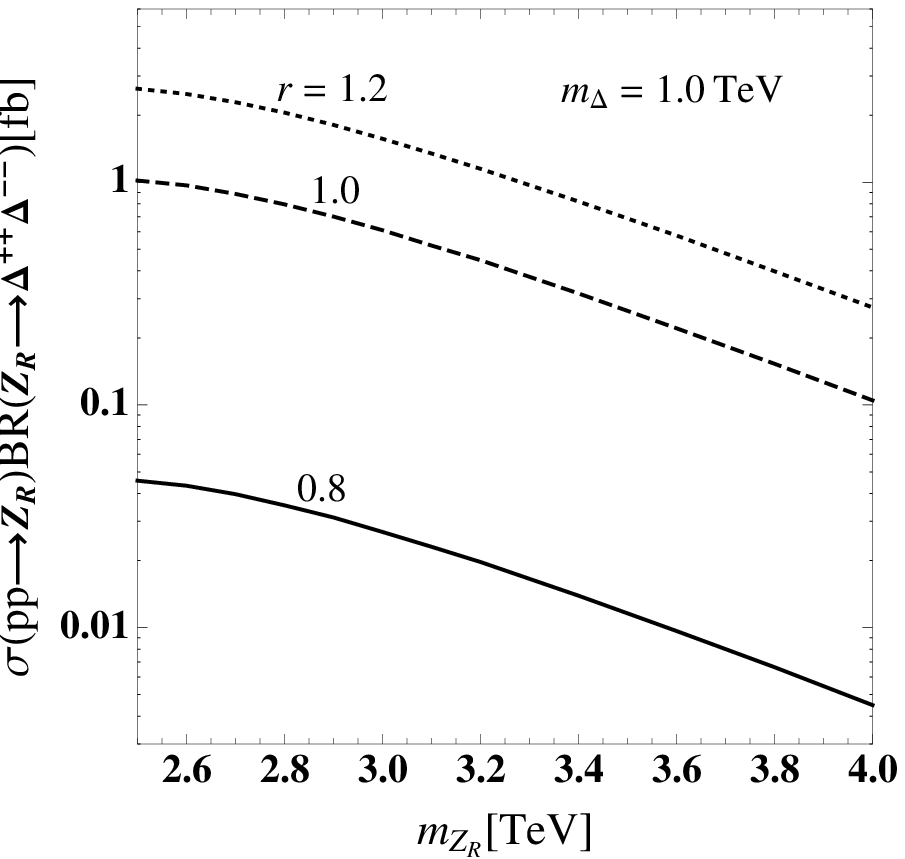}
   \caption{The left and right plots show $\sigma (pp \to Z_R) BR(Z_R \to \ell^+ \ell^-)$ and $\sigma (pp \to Z_R) BR(Z_R \to \Delta^{++} \Delta^{--})$ as a function of $Z_R$ mass for several values of $r \equiv g_R/g_L$ where we fixed doubly charged Higgs mass $m_{\Delta}$ as 1 TeV. The red curve in the left plot indicate the upper limit from LHC experiment~\cite{Aaboud:2016cth}.}
   \label{fig:CX}
\end{center}
\end{figure}
%%%%%%%%%%%%%%%%%%%%%%
Here we discuss the signature of our model at the LHC 13 TeV.
Then we focus on the doubly charged Higgs boson $\Delta^{\pm\pm}$, which decays into the same sign lepton pair with right-handed chirality. 
Particularly the process $pp \to Z_R \to \Delta^{++} \Delta^{--}$ is interesting since it provides clear four lepton signal where invariant masses of same sign leptons and of four leptons respectvely give mass of $\Delta^{\pm \pm}$ and $m_{Z_R}$~\footnote{The doubly charged Higgs pair can be produced via $\gamma$ and $Z$ exchange in s-channel. In this paper, we don't discuss these production processes since they are small as $< 0.1$ fb for TeV scale doubly charged Higgs and signal is less significant due to absence of peak in invariant mass of four leptons.}.
This is unlikely to neither the type II seesaw scenario nor the Zee-Babu type case with $k^{++}e^c_R e_R$, because the type II decay mode comes from the left-handed chirality, and the Zee-Babu type doubly charged Higgs is produced via gauge interaction with only $U(1)_Y$. Furthermore each of the component $y_{\Delta_R}$ can be determined through the neutrino oscillation data, CLFVs processes, and DM phenomenology such as $X$-ray line search. Thus we expect that collider signature further test the structure of the Yukawa coupling.

The gauge interactions associated with $Z_R$ are written as~\cite{Duka:1999uc, Ko:2015uma} 
\begin{equation}
{  \cal L} \supset \bar f_{SM} \gamma_\mu \left[ g_R c_M \left( Q - \frac{Q_{B-L}}{2 c_M^2} \right) Z_R^\mu \right] f_{SM} 
- \frac{Q_{B-L}}{2} \frac{s_M^2}{c_M} g_R Z_R^\mu (\Delta^{++} \partial_\mu \Delta^{--} - \Delta^{--} \partial_\mu \Delta^{++})
\end{equation}
where $Q$ is electric charge, $Q_{B-L}$ is $U(1)_{B-L}$ charge, $c_M \equiv \cos \theta_M = \tan \theta_W g_L/g_R$, $s_M \equiv \sin \theta_M$, and $g_R$ is $SU(2)_R$ gauge coupling.
Then we estimate the production cross section of $Z_R$ and its branching ratio (BR) with CalcHEP~\cite{Belyaev:2012qa} implementing the interaction and applying {\tt CTEQ6L} PDF~\cite{Nadolsky:2008zw}.
In Fig.~\ref{fig:CX}, we show $\sigma (pp \to Z_R) BR(Z_R \to \ell^+ \ell^-)$ and $\sigma (pp \to Z_R) BR(Z_R \to \Delta^{++} \Delta^{--})$ as a function of $m_{Z_R}$ for several values of $r \equiv g_R/g_L$
with fixed doubly charged Higgs mass $m_\Delta = 1$ TeV where constraint on  $\sigma (pp \to Z_R) BR(Z_R \to \ell^+ \ell^-)$ from LHC experiment is indicated by red curve~\cite{Aaboud:2016cth}.
We find that $Z_R$ should be heavier than around $3.5$ TeV where the lower mass limit depends on $r$. 
Note here that this result does not depend on doubly charged Higgs mass strongly if it is lighter than $m_{Z_R}/2$ sufficiently.
The doubly charged Higgs pair production cross section via $Z_R$ is given as $\sim \{0.28, 0.11, 0.045\}$[fb] for $r = \{1.2, 1.0, 0.8\}$ with $m_{Z_R} = 4$ TeV.
Thus $O(10)-O(100)$ number of events can be obtained with luminosity of $\sim 100-300$ fb$^{-1}$ for $r \geq 1$ while number of events is smaller for $r< 1$.
Therefore we can test our model at the LHC with sufficient luminosity since the four lepton final state gives clear signal,
and structure of the Yukawa coupling $y_{\Delta_R}$ would be investigated by measuring the BR of $\Delta^{\pm \pm}$.
The detailed simulation analysis including SM background is beyond the scope of our paper and it will be investigated in future work.

\subsection{Numerical analysis}
Now that all the formulae have been provided, we carry out numerical analysis to search for parameter region satisfying all the constraints.
First of all, we fix the following parameters in the scalar sector:
\begin{eqnarray}
m_{h_1}=125 \ {\rm GeV}, v_R{(\approx v_\Delta)}=10^5 \ {\rm GeV}, m_{\nu_{R,1}}=7.1 \ {\rm keV}. 
\end{eqnarray}
{
Before discussing the numerical analysis, we comment on the neutrinoless double beta decay.
Once we apply these above values, we can estimate the the neutrinoless double beta decay as
\begin{align}
m_{\beta\beta}\approx3.5\ {\rm meV},
\end{align}
where non-standatrd contribution is about ${\cal O}(10^{-5})$ meV.
It suggests that it satisfies the experimental bound on $m_{\beta\beta}\lesssim0.29$ eV, as discussed before.
}
Then we have 26 free parameters (see Appendix A) and randomly select the values of 
these parameters within the following ranges:  
\begin{eqnarray}
&M_{E, i} =\left(500 \ {\rm GeV}, 1000\ {\rm GeV} \right), 
m_{\nu_R, j} =\left(5000 \ {\rm GeV}, 10000 \ {\rm GeV}\right), 
m_{h_2}=\left(1000 \ {\rm GeV}, 10000 \ {\rm GeV}\right), \nn\\
&F_{L_{23}} = \left(0, 0.01\right), 
h_{L_{11}} =  \left(0, 0.01\right), 
h_{L_{12}} =  \left(0, 0.01\right), 
h_{L_{13}} =  \left(0, 1\right),  \nn\\
&F_{R_{23}} = \left(-0.01, 0\right), 
h_{R_{11}} =  \left(-0.01, 0\right), 
h_{R_{12}} =  \left(-0.01, 0\right), 
h_{R_{13}} =  \left(-1, 0\right), \nn\\
&\alpha_1 =  \left(-0.3, -0.2\right), 
\alpha_2 =  \left(0.2, 0.3\right), 
\alpha_3 =  \left(-0.0002, -0.00002\right), \nn\\
&y_{\Delta}^i = \left(-0.1, 0.1\right), 
\alpha_{R_1} = \left(2.9, \pi\right), 
\alpha_{R_2}=  \left(1.5, 2\right), 
\alpha_{R_3} =  \left(0.1, 0.5\right), \nn\\
&\alpha_I =  \left(-3, -2\right), 
\alpha_p = \left(0.05, 0.1\right), 
%\alpha_{pp} =\left(2.5, 3\right), 
\end{eqnarray}
which are found as preferred parameter range to satisfy the constraints.
%%%%%%%%%%%%%%%%%%
\begin{figure}[t]
\begin{center}
\includegraphics[width=0.80\columnwidth]{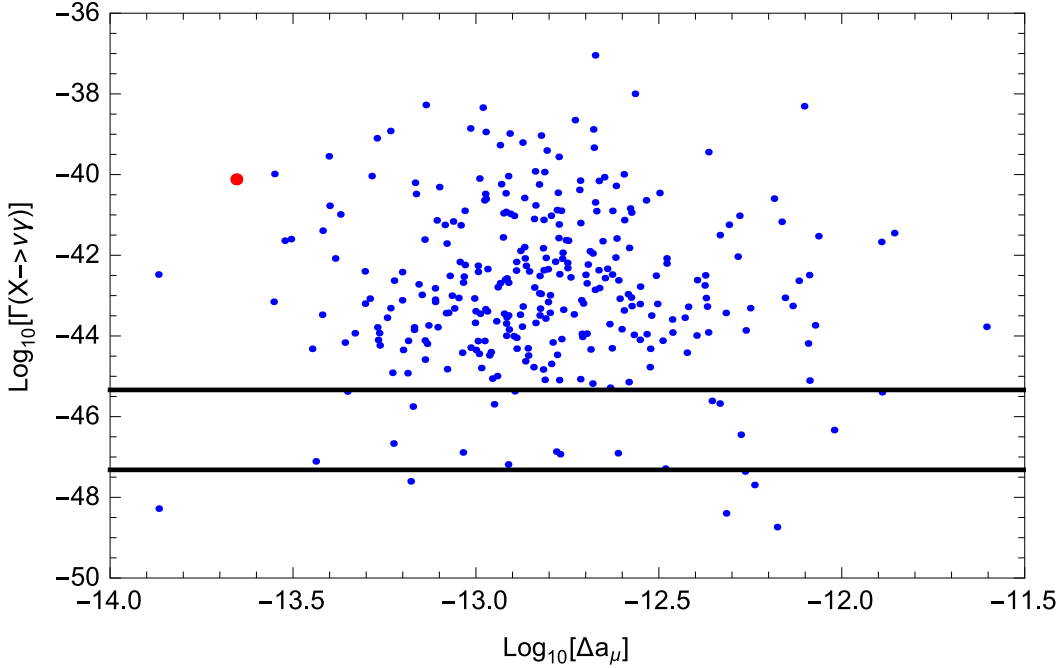}
   \caption{The correlation between $\Delta a_{\mu}$ and
 $\frac{\Gamma(X\to\nu_k\gamma)}{M_X}$. 
All points satisfy the current LFV constraints, {a red point means 
$\Delta a_{\mu}$ is negative and blue points mean $\Delta a_{\mu}$ is positive.}
The total number of the consistent data points is $311$.}
   \label{fig:scandata}
\end{center}
\end{figure}
%%%%%%%%%%%%%%%%%%%%%%
Then we have examined $10^6$ sampling points to investigate how much parameter space is allowed.  
{We find $311$ points that satisfy the current LFV constraints and the neutrino oscillation data.
Fig.~\ref{fig:scandata} shows the correlation between $\Delta a_{\mu}$ and $\frac{\Gamma(X\to\nu_k\gamma)}{M_X}$ where a red point represents 
negative $\Delta a_{\mu}$ and the blue points represent positive $\Delta a_{\mu}$. }
The DM decay rate can satisfy the experimental data {(16 points)} if the points are within the line between the two black horizontal lines in Fig.~\ref{fig:scandata}.
On the other hand,  the discrepancy of muon $g-2$ from SM is at most the order {${10^{-11}}$}, which is  too small to explain the experimental data by the order {$0.01$} magnitude. 
This is because there exist more negative contributions in {Eqs.~(III.5-7) than the positive contribution in Eqs.~(III.4).}
But a future experiment might verify the scale.

\section{Conclusions}
{We have studied a radiative seesaw model based on a $SU(2)_L \times SU(2)_R \times U(1)_{B-L}$ symmetry, where the neutrino mass matrix is induced at two loop level.
Then we have formulated masses in lepton sector, lepton flavor violating decay ratios, muon $g-2$, and the decay rate of the long lived dark matter. }
Due to the antisymmetric Yukawa couplings contributing to the active neutrino mass and {absence of $\Delta_L$}, a zero mass eigenstate (with two massive) is predicted, {and (a long-lived) dark matter candidate can be accommodated in our model}. 
{Then we have carried out a numerical  analysis to search for the parameter space which is consistent with all the experimental constraints, and correlation between $\Delta a_\mu$ and $\Gamma(X \to \nu_k \gamma)/M_X$ for the allowed parameter set is depicted in Fig.~\ref{fig:scandata},
in which the red points represent negative $\Delta a_{\mu}$ and the blue points represent positive $\Delta a_{\mu}$. }
The DM decay rate satisfies the experimental data {(16 points)} if the points are within the line between the two black horizontal lines.
On the other hand,  the discrepancy of muon $g-2$ from SM is at most the order {${10^{-11}}$} that is too small to explain the current experimental data by the order {$0.01$} magnitude, since there exist only a positive contribution comparing to three negative contributions. But a future experiment might verify the scale.

%Here the doubly charged boson does not play a role in our main analysis except for muon $g-2$, since we have assumed that  
%the Zee-Babu type neutrino mass that mediates the doubly charged boson is negligible.
% the canonical seesaw with one-loop induced Dirac mass is considered as the dominant contribution. 
%Although this assumption is appropriate as far as all the loop factors are the same order and there exists rather large hierarchy among  
% mediating fields, this contribution can still be relevant in a different situation such as absence of the isospin triplet boson $\Delta_R$.
%However this formulation is very difficult because all of the block mass matrix is rank 2 matrix.

Our model also could be tested at collider experiments by searching for exotic charged particles such as heavy leptons and doubly/singly charged Higgs bosons.
These particles would be produced via electroweak interactions at the LHC when their masses are $O(100)$ GeV to $O(1)$ TeV.
%Furthermore, since there are two doubly charged Higgs bosons we would distinguish our model from the other left-right symmetry models by searching for the signal of these doubly charged Higgs.
{Then we have analyzed doubly charged Higgs production via the process $pp \to Z_R \to \Delta^{++} \Delta^{--}$ at the LHC where doubly charged Higgs decays into two same sign leptons providing clear  signals from four lepton final states. The production cross section is estimated as $\sim 0.05$-$0.3$ fb for $m_{Z_R} = 4$ TeV depending on value of the ratio of $SU(2)_{L(R)}$ gauge couplings, and we can obtain around 10 to 100 number of events with luminosity of $O(100)$ fb$^{-1}$. Thus we can test the signature of our model and structure of Yukawa coupling for $\Delta_R$ and right-handed charged leptons could be tested by measuring branching ratio of doubly charged Higgs. }
The detailed analysis including SM background will be left as future works.

%\newpage
%%%%%%%%%%%%%%%%%%%%%%%%%%%%%%%%%%%
%\vspace{0.5cm}
%\hspace{0.2cm} {\bf Acknowledgments}
\section*{Acknowledgments}
\vspace{0.5cm}
Authors thank to Dr. Kei Yagyu for fruitful discussions.
H.O. expresses his sincere gratitude toward all the KIAS members, Korean cordial persons, foods, culture, weather, and all the other things.
This work was supported by the Korea Neutrino Research Center which is established by the National Research Foundation of Korea(NRF) grant funded by the Korea government(MSIP) (No. 2009-0083526) (Y.O.).
%%%%%%%%%%%%%%%%%%%%%%%%%%%%%%%%%%%

\begin{appendix}
\section{Yukawa couplings}
In this section, we discuss the structure of Yukawa couplings of our model.  
Our neutrino masses are obtained by $(\mathcal{M}_\nu)_{ab}\approx m_D m_{\nu_R}^{-1} m_D^T$,
where $m_{\nu_R}=y_{\Delta_R} v_\Delta$ and $m_D$ is given by Eq.(\ref{eq:mD}). 
Using the Casas-Ibarra parametrization, the $m_D$ is written by 
\begin{eqnarray}
m_D=U^*_{MNS}. diag \left( m_1^{\frac12},m_2^{\frac12},m_3^{\frac12}\right).O.m_{\nu_R}^{\frac12}, 
\end{eqnarray}
where $U_{MNS}$ is the MNS matrix, $m_i$'s are neutrino masses, $O$ is an complex orthogonal matrix.  
$O$ is parametrized by three complex parameters: $\alpha_1$, $\alpha_2$, $\alpha_3$. 

Generally, a matrix $M$ is factorized by the following form, 
\begin{eqnarray}
M=LDU, 
\end{eqnarray}
where $D$ is diagonal matrix and L (U) is upper (lower) triangular matrix with unit diagonal components. 
The factorization is called LDU decomposition. 
We can factorize $m_D$ using the LDU decomposition as follows: 
\begin{eqnarray}
m_D=LDU=L_D U_D, 
\end{eqnarray}
where $L_D\equiv LD^\frac12=F_L h_L Z_D^\frac12$, $U_D\equiv {D^\frac12U}=(F_R h_R Z_D^\frac12)^T$ 
and the diagonal matrix $Z_D$ is written by 
\begin{eqnarray}
Z_{D_{ii}}=\frac{v_L v_R \ln Z_{i}}{2 \pi^2 M_{E_i} ( 1-Z_{i} )}.
\end{eqnarray}
We assume $F_L h_L$ and $F_R h_R$ are lower triangular matrices.   
The components of $F_{L(R)}$ and $h_{L(R)}$ are written by the following form, 
\begin{eqnarray}
h_{L(R)_{31}}&=& l_{L(R)_{21}} - \frac{l_{L(R)_{22}}}{l_{L(R)_{32}}} 
\frac{l_{L(R)_{11}} l_{L(R)_{32}}}{l_{L(R)_{31}} l_{L(R)_{22}} - l_{L(R)_{21}} l_{L(R)_{32}}} h_{L(R)_{11}}, \nn\\
h_{L(R)_{21}}&=& -l_{L(R)_{31}} +\frac{l_{L(R)_{11}} l_{L(R)_{32}}}{l_{L(R)_{31}} l_{L(R)_{22}} - l_{L(R)_{21}} l_{L(R)_{32}}} h_{L(R)_{11}}, \nn\\
h_{L(R)_{32}}&=& l_{L(R)_{22}} - \frac{l_{L(R)_{22}}}{l_{L(R)_{32}}} 
\frac{l_{L(R)_{11}} l_{L(R)_{32}}}{l_{L(R)_{31}} l_{L(R)_{22}} - l_{L(R)_{21}} l_{L(R)_{32}}} h_{L(R)_{12}}, \nn\\
h_{L(R)_{22}}&=& -l_{L(R)_{32}} + \frac{l_{L(R)_{11}} l_{L(R)_{32}}}{l_{L(R)_{31}} l_{L(R)_{22}} - l_{L(R)_{21}} l_{L(R)_{32}}} h_{L(R)_{12}}, \nn\\
h_{L(R)_{23}}&=& \frac{l_{L(R)_{11}} l_{L(R)_{32}}}{l_{L(R)_{31}} l_{L(R)_{22}} - l_{L(R)_{21}} l_{L(R)_{32}}} h_{L(R)_{13}}, \quad%\nn\\
h_{L(R)_{33}}=\frac{h_{L(R)_{23}}h_{L(R)_{32}}}{h_{L(R)_{22}}}, \nn\\
F_{L(R)_{12}}&=&F_{L(R)_{23}}\frac{h_{L(R)_{23}} h_{L(R)_{32}}}{h_{L(R)_{13}} h_{L(R)_{22}}}, \quad %\nn\\
F_{L(R)_{13}}=- F_{L(R)_{23}}\frac{h_{L(R)_{23}} }{h_{L(R)_{13}}}. 
\label{yukawa}
\end{eqnarray}
In this case, $F_{L(R)} h_{L(R)}$ becomes a lower triangular matrix: 
\begin{eqnarray}
F_{L(R)} h_{L(R)}&=&\left( 
\begin{array}{ccc}
l_{L(R)_{11}} & 0 & 0 \\
l_{L(R)_{21}} & l_{L(R)_{22}} & 0 \\
l_{L(R)_{31}} & l_{L(R)_{32}} & l_{L(R)_{33}} \\
\end{array} 
\right)%\nn\\&=&
=L_D \  \text{or} \  U_D^T. 
\end{eqnarray} 
$l_{L(R)_{ij}}$ are determined by neutrino oscillation experiments. 
Therefore we have 8 free parameters: $h_{L(R)_{11}}$, $h_{L(R)_{12}}$, $h_{L(R)_{13}}$ and $F_{L(R)_{23}}$. 

\section{Loop factors for $\ell_i^- \to \ell_j^- \ell_k^+ \ell_\ell^-$}

Here we summarize the loop factors appearing in the formula of lepton flavor violating decay $\ell_i^- \to \ell_j^- \ell_k^+ \ell_\ell^-$ in Eq.~(\ref{eq:lfv-itojkl}).
\begin{align}
&A=\frac{-i}{2(4\pi)^2} \int dX 
\sum_{\alpha=1}^3 \sum_{a=1}^{2} 
\frac{ (H^b)_{\ell\alpha}(H^\dag_a)_{\alpha i} (H^a)_{j\beta}(H^\dag_b)_{\beta k} 
-
(H^b)_{\ell\alpha}(H^\dag_a)_{\alpha i} (H^b)_{j\beta}(H^\dag_a)_{\beta k}}
{x_1 M^2_{E_\alpha} +x_2 M^2_{E_\beta} +x_3 m^2_{h_a} +x_4 m^2_{h_b}}
%\nn\\
%&\hspace{1.2cm}
%-\frac{2i(ff^\dag)_{i\ell} (ff^\dag)_{jk} }{(4\pi)^2m^2_{h^\pm}}
,\\
&B=\frac{-i}{2(4\pi)^2} \int dX 
\sum_{\alpha=1}^3 \sum_{a=1}^{2} 
\frac{ (H^a)_{\ell\beta}(H^\dag_b)_{\beta k} (H^b)_{j\alpha}(H^\dag_a)_{\alpha i} 
-
(H^b)_{\ell\beta}(H^\dag_a)_{\beta k} (H^b)_{j\alpha}(H^\dag_a)_{\alpha i} }
{x_1 M^2_{E_\alpha} +x_2 M^2_{E_\beta} +x_3 m^2_{h_a} +x_4 m^2_{h_b}}%\nn\\
%&\hspace{1.2cm}
%+\frac{2i(ff^\dag)_{ij} (ff^\dag)_{k\ell} }{(4\pi)^2m^2_{h^\pm}}
,\\
&C=\frac{i}{(4\pi)^2} \int dX 
\sum_{\alpha=1}^3 \sum_{a=1}^{2} 
\frac{ M_{E_\alpha} M_{E_\beta} \left((H^b)_{\ell\alpha}(H^\dag_a)_{\alpha i} (H^a)_{j\beta}(H^\dag_b)_{\beta k} +
(H^b)_{\ell\alpha}(H^\dag_a)_{\alpha i} (H^b)_{j\beta}(H^\dag_a)_{\beta k}\right)}
{(x_1 M^2_{E_\alpha} +x_2 M^2_{E_\beta} +x_3 m^2_{h_a} +x_4 m^2_{h_b})^2},\\
&D=\frac{i  }{(4\pi)^2} \int dX 
\sum_{\alpha=1}^3 \sum_{a=1}^{2} 
\frac{ M_{E_\alpha} M_{E_\beta}
\left( (H^a)_{\ell\beta}(H^\dag_b)_{\beta k} (H^b)_{j\alpha}(H^\dag_a)_{\alpha i} +
(H^b)_{\ell\beta}(H^\dag_a)_{\beta k} (H^b)_{j\alpha}(H^\dag_a)_{\alpha i} \right)}
{(x_1 M^2_{E_\alpha} +x_2 M^2_{E_\beta} +x_3 m^2_{h_a} +x_4 m^2_{h_b})^2},
\end{align}
\begin{align}
&A_L=\frac{4i(f_L f_L^\dag)_{ij} (f_L f_L^\dag)_{k\ell} }{(4\pi)^2 m^2_{h^\pm}}
%\nn\\&\hspace{1.2cm}-
%%%
-\frac{i }{2(4\pi)^2} \int dX 
\sum_{\alpha=1}^3 \sum_{a=1}^{2} 
\frac{M_{E_\alpha} M_{E_\beta} \left( d + h (\ell \leftrightarrow j)\right) }
{(x_1 M^2_{E_\alpha} +x_2 M^2_{E_\beta} +x_3 m^2_{h_a} +x_4 m^2_{A_b})^2},\\
%%%
& B_L=-\frac{4i(f_L f_L^\dag)_{i\ell} (f_L f_L^\dag)_{kj} }{(4\pi)^2 m^2_{h^\pm}},\\
&A_R= \frac{-i}{2(4\pi)^2} \int dX 
\sum_{\alpha=1}^3 \sum_{a=1}^{2} 
\frac{ (H^{'a})_{\ell\beta}(H^{'\dag}_b)_{\beta k} (H^{'b})_{j\alpha}(H^{'\dag}_a)_{\alpha i} 
-
(H^{'b})_{\ell\beta}(H^{'\dag}_a)_{\beta k} (H^{'b})_{j\alpha}(H^{'\dag}_a)_{\alpha i} }
{x_1 M^2_{E_\alpha} +x_2 M^2_{E_\beta} +x_3 m^2_{A_a} +x_4 m^2_{A_b}}\nn\\
&\hspace{1.2cm}+
%%%
\frac{i}{2(4\pi)^2} \int dX 
\sum_{\alpha=1}^3 \sum_{a=1}^{2} 
\frac{ e+f-g-h }
{x_1 M^2_{E_\alpha} +x_2 M^2_{E_\beta} +x_3 m^2_{h_a} +x_4 m^2_{A_b}}\nn\\
&\hspace{1.2cm}-
%%%
\frac{i }{2(4\pi)^2} \int dX 
\sum_{\alpha=1}^3 \sum_{a=1}^{2} 
\frac{M_{E_\alpha} M_{E_\beta} \left( c + g(\ell \leftrightarrow j)\right) }
{(x_1 M^2_{E_\alpha} +x_2 M^2_{E_\beta} +x_3 m^2_{h_a} +x_4 m^2_{A_b})^2}\nn\\
&\hspace{1.2cm}
+ 8i(f_R)_{a\ell}(f_R)_{a'j} (f_R^\dag)_{ka} (f_R^\dag)_{ia'} J_{1,aa'},\\
&B_R =\frac{-i}{2(4\pi)^2} \int dX 
\sum_{\alpha=1}^3 \sum_{a=1}^{2} 
\frac{ (H^{'b})_{\ell\alpha}(H^{'\dag}_a)_{\alpha i} (H^{'a})_{j\beta}(H^{'\dag}_b)_{\beta k} 
-
(H^{'b})_{\ell\alpha}(H^{'\dag}_a)_{\alpha i} (H^{'b})_{j\beta}(H^{'\dag}_a)_{\beta k}}
{x_1 M^2_{E_\alpha} +x_2 M^2_{E_\beta} +x_3 m^2_{A_a} +x_4 m^2_{A_b}}\nn\\
&\hspace{1.2cm}+
%%%
\frac{i}{2(4\pi)^2} \int dX 
\sum_{\alpha=1}^3 \sum_{a=1}^{2} 
\frac{ a + b -c - d}
{x_1 M^2_{E_\alpha} +x_2 M^2_{E_\beta} +x_3 m^2_{h_a} +x_4 m^2_{A_b}}\nn\\
&\hspace{1.2cm}-8i(f_R)_{a\ell}(f_R)_{a'j} (f_R^\dag)_{ka'} (f_R^\dag)_{ia} J_{1,aa'},\\
&C_R=16i m_{\nu_{R,a}} m_{\nu_{R,a'}}  (f_R)_{a\ell}    (f_R)_{aj} (f_R^\dag)_{ka'}   (f_R^\dag)_{ia'}  J_{2,aa'}, 
\end{align}
\begin{align}
& E = \frac{i  }{2(4\pi)^2} \int dX 
\sum_{\alpha=1}^3 \sum_{a=1}^{2} 
\frac{ M_{E_\alpha} M_{E_\beta}\left(a + e(\ell \leftrightarrow j) \right)}
{(x_1 M^2_{E_\alpha} +x_2 M^2_{E_\beta} +x_3 m^2_{h_a} +x_4 m^2_{A_b})^2}, \\
& F = \frac{i  }{2(4\pi)^2} \int dX 
\sum_{\alpha=1}^3 \sum_{a=1}^{2} 
\frac{M_{E_\alpha} M_{E_\beta}\left( b + f(\ell \leftrightarrow j) \right)}
{(x_1 M^2_{E_\alpha} +x_2 M^2_{E_\beta} +x_3 m^2_{h_a} +x_4 m^2_{A_b})^2}.
\end{align}
%%%
Here the factors $\{a,b,..,h\}$ have been defined as
\begin{align}
&a = (H^{'b})_{\ell\alpha}(H^{\dag}_a)_{\alpha i} (H^{a})_{j\beta}(H^{'\dag}_b)_{\beta k} ,\quad
b = (H^{a})_{\ell\alpha}(H^{'\dag}_b)_{\alpha i} (H^{'b})_{j\beta}(H^{\dag}_a)_{\beta k} ,\\
%%%
&c = (H^{'b})_{\ell\alpha}(H^{\dag}_a)_{\alpha i} (H^{'b})_{j\beta}(H^{\dag}_a)_{\beta k} ,\quad
d = (H^{a})_{\ell\alpha}(H^{'\dag}_b)_{\alpha i} (H^{a})_{j\beta}(H^{'\dag}_b)_{\beta k} ,\\
%%% %%%
&e = (H^{a})_{\ell\beta}(H^{'\dag}_b)_{\beta k} (H^{'b})_{j\alpha}(H^{\dag}_a)_{\alpha i} ,\quad
f =  (H^{'b})_{\ell\beta}(H^{\dag}_a)_{\beta k} (H^{a})_{j\alpha}(H^{'\dag}_b)_{\alpha i}  ,\\
%%%
&g = (H^{'b})_{\ell\beta}(H^{\dag}_a)_{\beta k} (H^{'b})_{j\alpha}(H^{\dag}_a)_{\alpha i} ,\quad
h =  (H^{a})_{\ell\beta}(H^{'\dag}_b)_{\beta k} (H^{a})_{j\alpha}(H^{'\dag}_b)_{\alpha i}  ,\\
%%%
&J_{i,aa'}\equiv \frac{1}{(4\pi)^2}\int_0^1dx \int_0^{1-x}dy\frac{1-x-y}{[x m^2_{\nu_{R,a}}+ym^2_{\nu_{R,a'}}+(1-x-y)m^2_{h^\pm}]^{i} },\quad
i=(1,2),
\end{align}
where we define coupling factors $H^a_{ij}\equiv \frac{(h_L)_{ij} (O_R^T)_{1a}+ (h_R)_{ij} (O_R^T)_{2a} }{2\sqrt2}$ and $H^{'a}_{ij}\equiv \frac{(O^T_I)_{1a}(h_R)_{ij}}{\sqrt2}$, 
$\int dX\equiv  \int_0^1dx_1 dx_2dx_3dx_4\delta(x_1+x_2+x_3+x_4-1)$.

\end{appendix}

\end{document}